\documentclass[prb,preprint]{revtex4-1} 

\usepackage{amsmath}  
\usepackage{amsfonts} 
\usepackage{graphicx} 

\begin{document}


\title{A Simple Setup for the Determination of the Cosmic Muon Magnetic Moment }

\author{Damir Bosnar}
\email{bosnar@phy.hr} 
\author{Mihael Makek} 
\affiliation{Department of Physics, Faculty of Science, University of Zagreb, Croatia}

\author{Zoran Mati\'{c}}
\affiliation{Atir d.o.o., Zagreb, Croatia}


\date{\today}

\begin{abstract}
We present a simple setup for the measurement of the magnetic moment of cosmic muons convenient for a student laboratory experiment. 
A significant simplification is made in the detector system compared to previous experiments of this kind, yet it retains all the necessary  functionality of the system. This simplification additionally allows for the use of low-cost custom-made electronics for data readout and storage.
These improvements considerably reduce the cost and provide a more accessible setup for a student laboratory experiment.
\end{abstract}

\maketitle 

\section{Introduction}

Measurements of the properties of cosmic muons provide excellent opportunities for educational experiments in elementary particle physics. 
A classic and very popular example is the muon lifetime measurement, an experiment that has been conducted in numerous variations using standard nuclear physics equipment (see, for example, a recent review given by Riggi \emph{et al.}\cite{riggi}) and also with more accessible low-cost electronics. \cite{coan, bosnar} More complex measurements of the muon mass have also more recently been performed. \cite{mass1, mass2}

Conceptually and instrumentally much more demanding is the measurement of the muon magnetic moment. Being a point-like charged lepton that carries a unit electrical charge, either
positive or negative, and with a spin of 1/2 and a mass of about 207 masses of the electron, the muon possesses a magnetic moment.  \cite{PDG}
One of the most effective and successful measurements of the magnetic moment of cosmic muons as an educational experiment using standard nuclear physics equipment  was performed by C. Amsler. \cite{amsler} This measurement exploits the non-conservation of parity in the production and decay chain of muons, first observed in the experiment of Garwin\emph{et al.}, \cite{garwin} which was also the incentive for Amsler's experiment. 
In particular, as a consequence of parity violation, there is a net polarization of cosmic muons that have been stopped in a material at Earth's surface. By putting this material in a known magnetic field, the spins of the stopped muons precess in the field. The spins of the  muons are correlated to the direction of the positrons  (electrons) emitted in the muon decays. By measuring the distribution of positrons (electrons) in a certain direction as a function of time, it is possible to determine the precession frequency of the muons and from this the 
magnetic moment of the muon.

Our measurement is based on the same idea, but the detector system is simplified to the bare minimum of only one scintillation detector. 
Instead of using standard nuclear and particle physics equipment, low-cost electronics is used for the processing and recording of the signals from the scintillation detector developed for a simple system for muon lifetime measurement. \cite{bosnar}

In Sec.~\ref{section2}, we briefly describe the origin of cosmic muons and their properties relevant for this measurement. In Sec.~\ref{section3}, we introduce the muon magnetic moment and elucidate the principle of the measurement. Our experimental setup and measurements are described in Sec.~\ref{section4}. Section~\ref{section5} presents the analysis of our acquired data and the obtained results, which are then discussed in Sec.~\ref{section6}.

\section{Cosmic muons: origin, polarization, and decay}\label{section2}
By the term \emph{cosmic muons} we mean muons which are produced in the decays of pions and kaons originating  from the interactions of primary cosmic rays (mainly protons and light nuclei) with the nuclei in the Earth's atmosphere; see, for example, the textbook of D. Perkins.\cite{perkins} 
Cosmic muons from pion decay are much more common because, in the primary interactions pions ( $\pi^+$, $\pi^0$, $\pi^-$) are created with a much greater probability than kaons ( $K^+$, $K^0$, $K^-$). While focused on pion decay, the subsequent discussion is also valid for muons from kaon decay.

Charged pions decay by weak interaction according to:
\begin{equation}
 \pi^+ \rightarrow \mu^+ + \nu_{\mu} \quad \textrm{and} \quad  \pi^-  
 \rightarrow \mu^- + \bar{\nu}_{\mu}.
\end{equation}

In the pion rest frame, decays are isotropic in the full solid angle of 4$\pi$.
However, since created pions have given momenta toward Earth, determined by the initial directions of primary cosmic rays,  most of the produced muons also have momenta directed toward the Earth.
On Earth's surface, the mean muon energy is approximately 4 GeV and the total downward-directed flux of muons with energies higher than 1 GeV/c is approximately 1 cm$^{-2}$ min$^{-1}$.  Due to the positive charge of the primary cosmic rays, 
the ratio of the number of positive-to-negative cosmic muons is about 1.25 in the GeV range. \cite{PDG}

In pion decay, there is maximal violation of parity and, in the rest frame of a positive pion, the produced positive muon is always left-handed (meaning its spin is in the opposite direction of its momentum). This property is the consequence of the conservation of the total angular momentum  and the fact that pions have spin zero and neutrinos can only be left-handed. On the other hand, since antineutrinos can 
only be right-handed (i.e., their spin is in the direction of their momentum), negative muons can only be right-handed when produced in the rest frame of the negative pions; see, for example, the textbook of D. Griffiths.\cite{griffiths}

Polarization of the cosmic muons is the property resulting from a preferred  orientation of their spin, see, for example, articles of S. Hayakawa \cite{haya} and S. J. Johnson. \cite{john} 
At ground level it is caused by parity violation in the processes of pion decay and by the energy spectra of pions from which the muons are produced. To explain this point, let us consider two extreme cases, one in which the muon is emitted in the direction of the pion boost ( i.e., toward Earth's surface) in the rest frame of the decaying pion, \emph{forward muon},  and the other in which the muon  is emitted in the opposite direction of the pion boost (i.e., away from Earth's surface) in the rest frame of the decaying pion, \emph{backward muon}. 

\begin{figure}[h]
{\includegraphics[width=100mm]{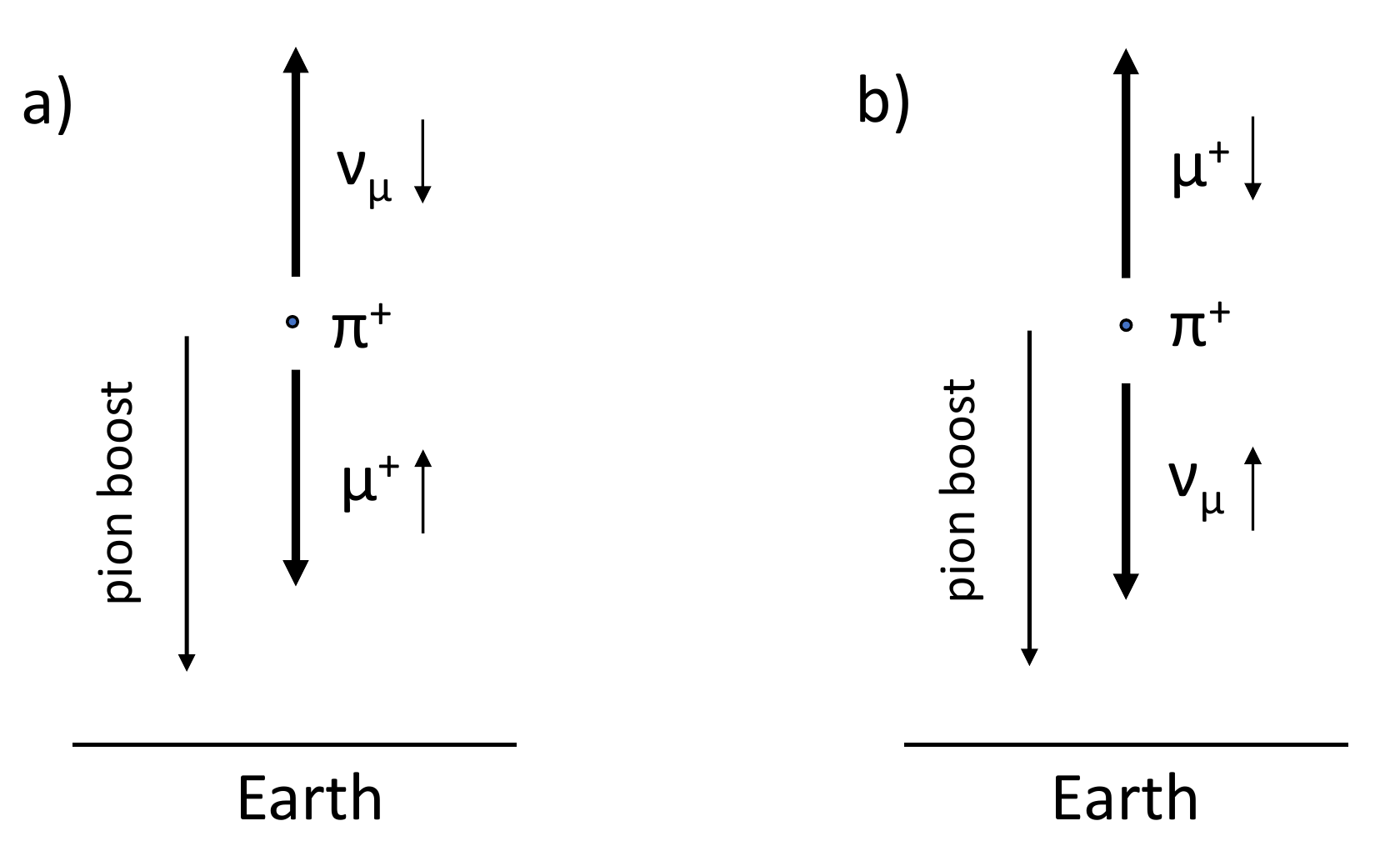}}
\caption{a) A positive muon emitted in the pion boost direction in the rest frame of decaying pion. b) A positive muon emitted in the opposite direction of the pion boost in the rest frame of a decaying pion.}
\label{pion-decay}
\end{figure}

If we consider positive pions, in the forward muon case, the $\mu^+$ is emitted toward Earth and the $\nu_{\mu}$ is emitted in the opposite 
direction, as in Fig.~\ref{pion-decay}(a). Thus, forward muons, which are initially left-handed, remain left-handed when they arrive at Earth's surface. In the backward muon case, the momentum of the produced $\mu^+$ in the rest frame of the decaying pion is in the opposite direction of the pion boost, but since it is left-handed, its spin is in the boost direction, that is, in the direction toward Earth, as in  Fig.~\ref{pion-decay}(b). Positive backward muons that reach Earth's surface (i.e., the muons for which the boosts of the decaying pions were sufficiently large to reverse the initial direction of the backward muons) will be right-handed at Earth's surface. 
Since the number of pions decreases with the pion energy, an excess number of forward muons will reach Earth's surface, producing a net upward polarization of positive muons at Earth's surface. 

The analogous consideration is also valid for negative muons and, keeping in mind that they are produced as right-handed particles, their net polarization at Earth's surface will be downward.

Muon polarization has been measured at Earth's surface as a function of energy and zenith angle. For example, for positive muons from the zenith, the polarization has been determined to be $-0.38 \pm 0.04$ and $-0.33 \pm 0.04$ for energies near 3.5  Ge and 1.0 GeV, respectively; 
the polarization falls off at lower muon energies. \cite{turner} The negative sign indicates that the spin is antiparallel to the direction of the flight of the positive muons. The observed polarization of the muons at Earth's surface indeed proves the parity violation in the pion decay. 

\begin{figure}[h]
{\includegraphics[width=100mm]{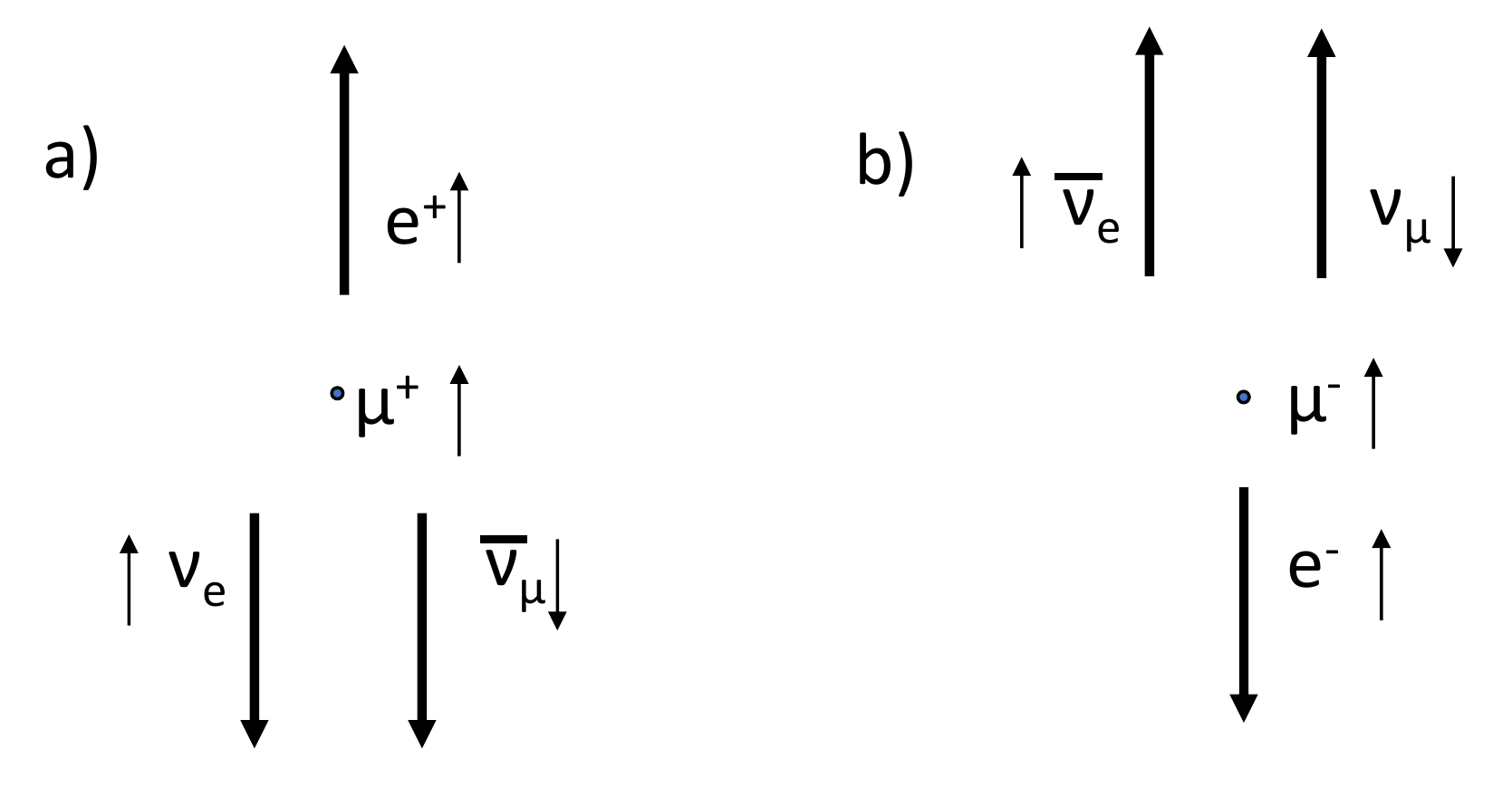}}
\caption{Decay at the rest of a positive  (a) and a negative (b) muon with the emission of the most energetic positron and electron, respectively.}
\label{muon-decay}
\end{figure}

Muons decay via the weak interaction with a mean lifetime of $\tau = 2.1969811 \pm 0.0000022 \,\mu$s.\cite{PDG} Parity violation is also present in these decay processes, which proceed as follows:
\begin{equation}
 \mu^+ \rightarrow e^+ +\nu_e + \bar{\nu}_{\mu} \quad \textrm{and} \quad 
 \mu^- \rightarrow e^- + \bar{\nu}_e + \nu_{\mu}.
\end{equation}
The directions of the emitted positrons (electrons) in these decays are correlated with the above described net polarization of the muons at Earth's surface, producing an asymmetric angular distribution. This fact can be qualitatively understood if,  in the decay of a muon at rest, we consider the extreme case where the most energetic positron (electron) is emitted. \cite{okun} In this case, both the neutrino and antineutrino must be emitted in the direction opposite that of the positron (electron), as in Fig.~\ref{muon-decay}. Since neutrinos and antineutrinos have an opposite handedness, the total angular momentum carried by them is zero, and a positron (electron) must be emitted with the spin parallel to that of the decaying muon. To meet this requirement, and since positrons (electrons) are right-handed (left-handed), their momenta should be directed in (opposite to) the directions of the muon spin. 
So, for the decays of positive muons, positrons are favorably  emitted in the  direction of the muon spin,  and for the  decays of negative muons, electrons are favorably emitted in the opposite direction of the muon spin.

\section{Muon magnetic moment and principle of measurement}\label{section3}

A charged particle with a spin carries a magnetic moment. For the muon, which is a point-like lepton and has charge $e$, mass $m_{\mu}$, and spin $\vec{s}$, the magnetic moment is given by
\begin{equation}
 \vec{\mu}_{\mu} = g_{\mu}\, \mu _{\mu}^{\rm{Bohr}} \, \vec{s}
\end{equation}
where $g_{\mu}$ is the proportionality factor called the gyromagnetic ratio and $\mu _B$ is the Bohr magneton of the muon given by $\mu_{\mu}^{\rm{Bohr}}=\frac{e\hbar}{2m_{\mu}}$. The gyromagnetic ratio is predicted by the Dirac equation to be exactly 2 for leptons, but the experimentally determined $g_{\mu}$-value differs slightly from this theoretical expectation, an inconsistency termed the \emph{magnetic moment anomaly},\cite{PDG} a subject of great interest in particle physics both experimentally \cite{mion-mag-exp} and theoretically. \cite{mion-mag-th}

If a muon is in an external magnetic field $\vec{B}$, its spin will precess with the frequency 
\begin{equation}
 \omega=g_{\mu}\,\frac{eB}{2m_{\mu}}.
 \label{eq:omega}
\end{equation}
 
If cosmic muons are stopped in a suitable non-magnetic material that is immersed in a horizontal magnetic field (with respect to Earth's surface), there will be a precession of  spins of the stopped muons. 
Copper is one such suitable material because it effectively stops cosmic muons due to its relatively high density and atomic number.
Most of the negative muons are quickly caught by nuclei in the stopping material (the lifetime of negative muons in copper is
0.163 $\mu$s \cite{eck}) and since our apparatus is not sensitive enough to determine their influence, in the following discussion we restrict ourselves to the positive muons, which stop in the material and decay as free particles. 

The muon spin precession is determined as follows. We detect positrons from the decays of stopped muons only in the upward direction. Above, we described the net polarization of the cosmic muons at Earth's surface, the precession of their spins in the magnetic field, and the correlations of the directions of emitted positrons with muon spins. Based on these facts, there will be an oscillation in time of the number of detected positrons from the decays about the usual exponential decay curve. When there is no magnetic field present, the data from our experiment should follow a simple exponential dependence
\begin{equation}
I(t)=I_0 e^{-\frac{t}{\tau}} + C
\label{eq:I1}
\end{equation}
where $I(t)$ is the number of muon decays in a unit of time at time $t$, $I_0$ is the decay rate at $t=0$, $\tau$ is the muon mean lifetime, and a constant $C$ due to the background. With an applied magnetic field, with good accuracy, we expect the above relation to become\cite{amsler}
\begin{equation}
I(t)=I_0 e^{-\frac{t}{\tau}}[1+\alpha \cos(\omega t + \delta)] + C,
\label{eq:I2}
\end{equation}
where $\alpha$ is the experimental asymmetry (that depends on the asymmetry of the positrons emitted in the muon decays, the polarization of the stopped muons, and the geometry of the setup), $\omega$ is the muon precession frequency, and $\delta$ is the angle of initial muon polarization.

By fitting function Eq.~(\ref{eq:I2}) to the measured time distribution, which is obtained by detecting upward emitted positrons from the decays of muons that are stopped in the material in the horizontal magnetic field, we can determine the frequency of the muon precession $\omega$, and hence $g_{\mu}$ from Eq.~(\ref{eq:omega}).

\section{Experimental setup and measurements}\label{section4}
We used one scintillation detector for the detection of the incoming cosmic muons and the same detector for the detection of the positrons from the decays. The scintillation detector consisted of an NE 102A scintillator plate of 50 cm $\times$ 50 cm $\times$ 1~ cm, coupled to a twisted light guide of 60 cm in length on which a Philips XP5312/SN 3-inch photomultiplier tube was attached. The scintillator and the light guide were glued with optical cement and wrapped with thin aluminum foil and with black scotch tape. The PMT was mounted in an aluminum tube with holders that enabled tight coupling to the light guide; the coupling was enhanced by optical grease. A box with a custom-made high-voltage supply with variable output voltage was attached to the aluminum tube that housed the PMT. The voltage applied in the measurement was $-1060$~ V. 

A block of  copper measuring 50 cm $\times$ 50 cm $\times$ 2.5 cm was placed beneath the scintillator plate. The copper block consisted of 40 pieces of 2.5 cm $\times$ 2.5 cm $\times$ 25 cm copper bars. The distance between the middle of the copper block and the middle of the scintillator plate was 6.3 cm. The role of the copper block was to stop some of the incoming cosmic muons. 

The copper block was positioned centrally inside a rectangular copper coil of one layer with 355 windings, a length of 60 cm, a width of 57.5 cm and a height of 7.5 cm, and whose purpose was to produce the necessary magnetic field. The wire in the coil had a cross section of 1.5 mm$^2$. The copper coil was mounted on an aluminum frame, which also held the copper bars and the scintillation detector. During the measurements with the currents, the coil was cooled by a fan. 

\begin{figure}[h]
{\includegraphics[width=80mm]{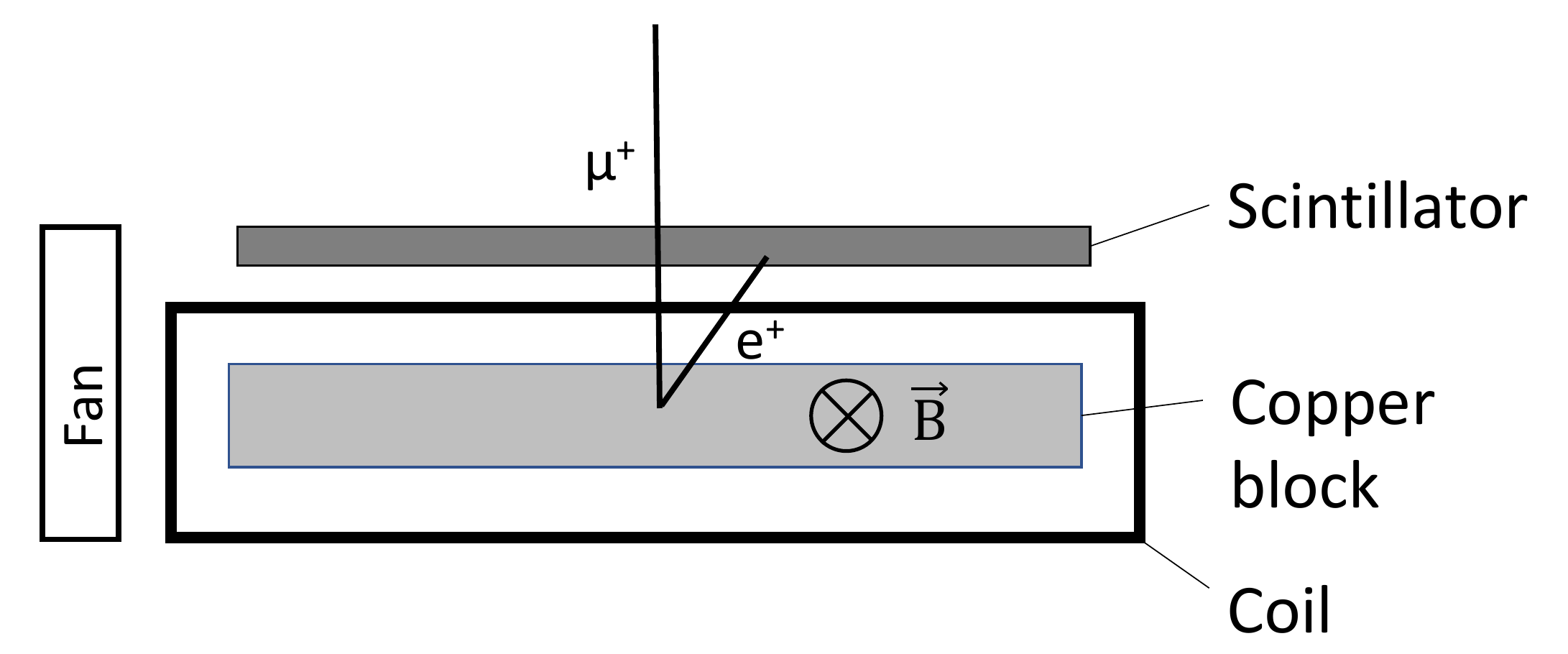}}
\caption{Schematic representation of the experimental setup (not to scale, dimensions are given in the text).}
\label{setup}
\end{figure}

The experimental setup is schematically shown in Fig.~\ref{setup}.

With a current of 5 A through the coil, we obtained a magnetic field of approximately 3.7 mT at the center of the coil. The field fell to approximately 3.4 mT at 25 cm from the coil's center (i.e., at a position toward the open edges of the coil and the end of the copper target). With a current of 6 A, the magnetic fields were $\sim$4.3 mT and  $\sim$3.9 mT at the coil's center and at 25 cm from the coil's center (towards the open edges of the coil), respectively. 
No significant variation was observed in the measured field (measured approximately every 10 cm) on the lines transverse to the longitudinal axis of the coil. These measurement were made at at several distances along the longitudinal axis from the center of the coil.

In the calculation of $g_{\mu}$ from Eq. (\ref{eq:omega}) and for its uncertainty estimation we used the average of the fields measured in the center of the coil and at 25 cm from the coil's center (towards the open edges of the coil) and the corresponding uncertainties.

Most of the cosmic muons come from above and the scintillation detector registered the muons that pass  through or stop and decay in the copper block beneath it.  The same scintillation detector also registered the positrons that were emitted in the upward direction in the decays of the stopped muons in the copper block.

\begin{figure}[h]
{\includegraphics[width=120mm]{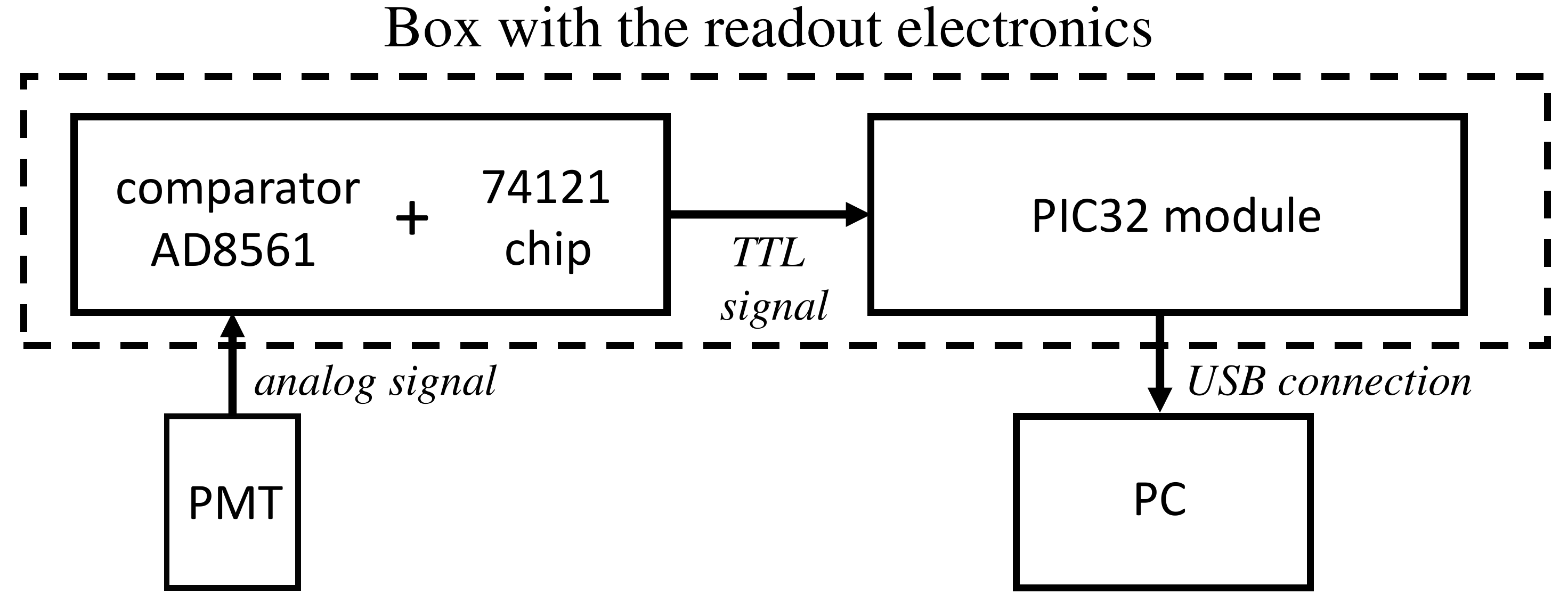}}
\caption{Schematic representation of the electronics for signal processing and data acquisition.}
\label{el}
\end{figure}

 For signal processing and data acquisition, we used custom-made low-cost electronics developed for the simple setup for the measurement of the cosmic muon lifetime.\cite{bosnar} Negative analog signals from the PMT of the scintillation detector were input to the fast TTL comparator (Analog Devices AD8561) with a variable threshold  in the range of 0 to $-500$~ mV. To secure a constant width of the output pulse, regardless of the input pulse width, a Texas Instruments 74121 chip was used.  Positive logic signals, produced by the signals from the scintillation detector above the set threshold, were input to the 32-bit microcontroller 
 (MikroElektronika PIC32 module, model MINI-32 BOARD). The PIC32 module recorded time stamps of the events with a time resolution of 12.5~ns, which were then stored on a PC disk. A schematic representation of the electronics is given in Fig.~\ref{el}.  

When started, the time span of the measurement by the PIC32 module was approximately 54 seconds ($2^{32}$ $\times$12.5 ns) and after this time it was reset automatically to zero, and the measurement continued automatically until it was stopped by the user.
The module was programmed to transfer the recorded data in a block of 16 events to a computer via serial/USB connection and to perform simple operational tasks on the data. It counted the total number of registered events during the measurement and the number of events for which the time difference was less than 10 $\mu$s, and these numbers were shown on a display. The last
number indicates the number of registered muon decays in the selected
time window, as explained in the following sections.

With this setup we performed one measurement of approximately 5 days without the magnetic field,
one measurement of 20 days with a current of 5 A through the copper wire, and one measurement of 21 days with a current of 6 A. 

\section{Data analysis and results}\label{section5}
Extractions of the time distributions of the muon decays were performed in an identical way as for the muon lifetime measurement.  \cite{bosnar} We first calculated the time differences between every pair of subsequently registered events. By selecting the time differences  smaller than some selected time window, e.g., 10 $\mu$s, we can extract events that, with high probability, correspond to the decays of the stopped muons. 
Since the time resolution of the PIC32 unit is pretty high, 12.5 ns per channel, even with the measurements
of several weeks, the number of events registered in each channel will be small. To obtain 
a statistically significant number of counts we selected time intervals in which we counted the number of muons that decay.

We used time intervals $\Delta T=0.2 \,\mu$s and fitted Eqs.~(\ref{eq:I1}) or (\ref{eq:I2}) on the obtained distributions for the measurement without a magnetic field or with a magnetic field, respectively.

\begin{figure}[h]
{\includegraphics[width=80mm]{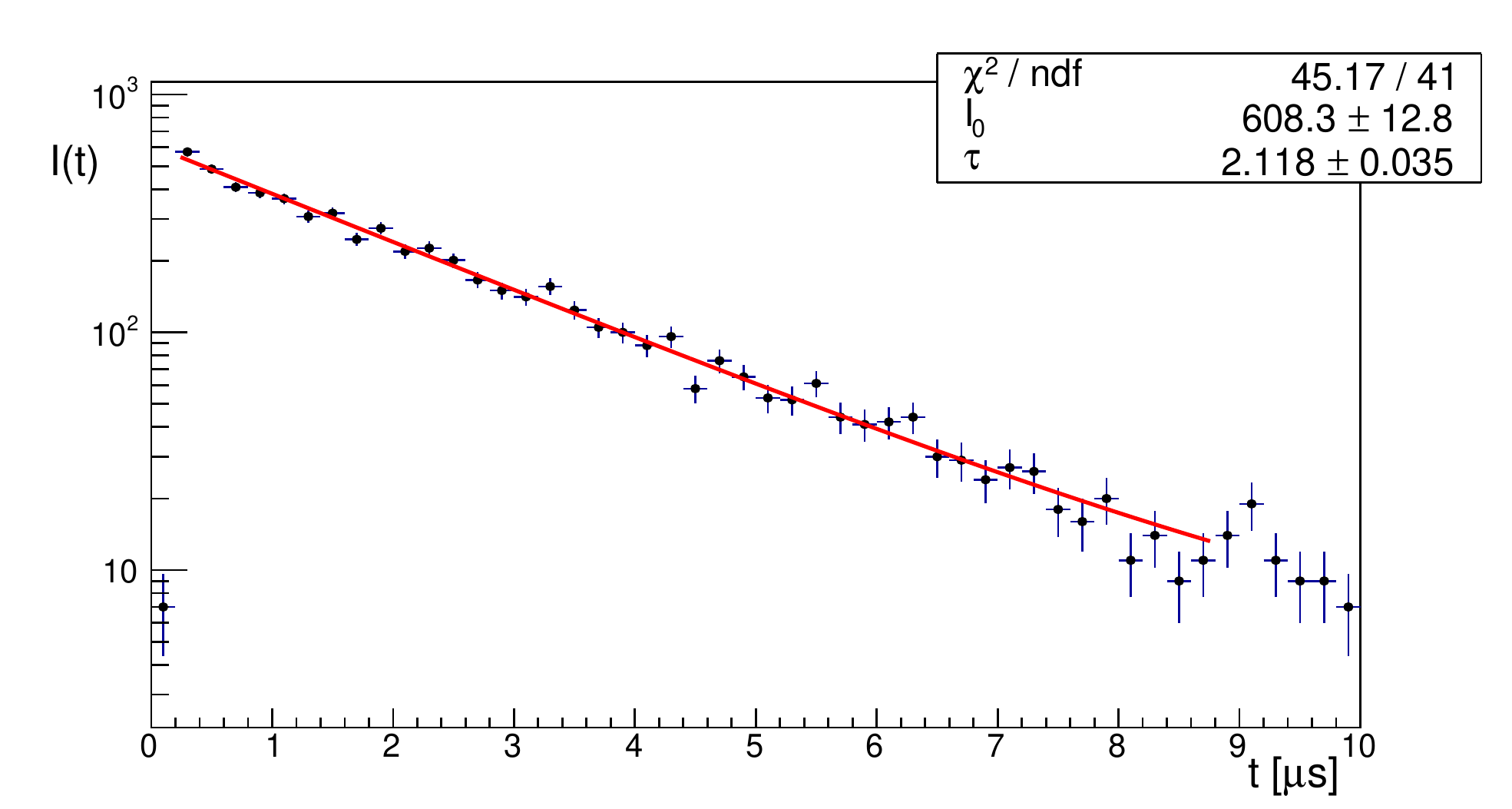}}
\caption{Measurement without a magnetic field and the fit to Eq.~(\ref{eq:I1}).}
\label{no-field}
\end{figure}

For the measurement without a magnetic field, we obtained 5981 events within the time window of 10 $\mu$s (a total of 2,942,766 events was registered with the scintillation detector).
The obtained time distribution and fit are shown in Fig.~\ref{no-field}. We obtained a muon mean life of $\tau$ = 2.12 $\pm$ 0.04 $\mu$s.

\begin{figure}[h]
{\includegraphics[width=80mm]{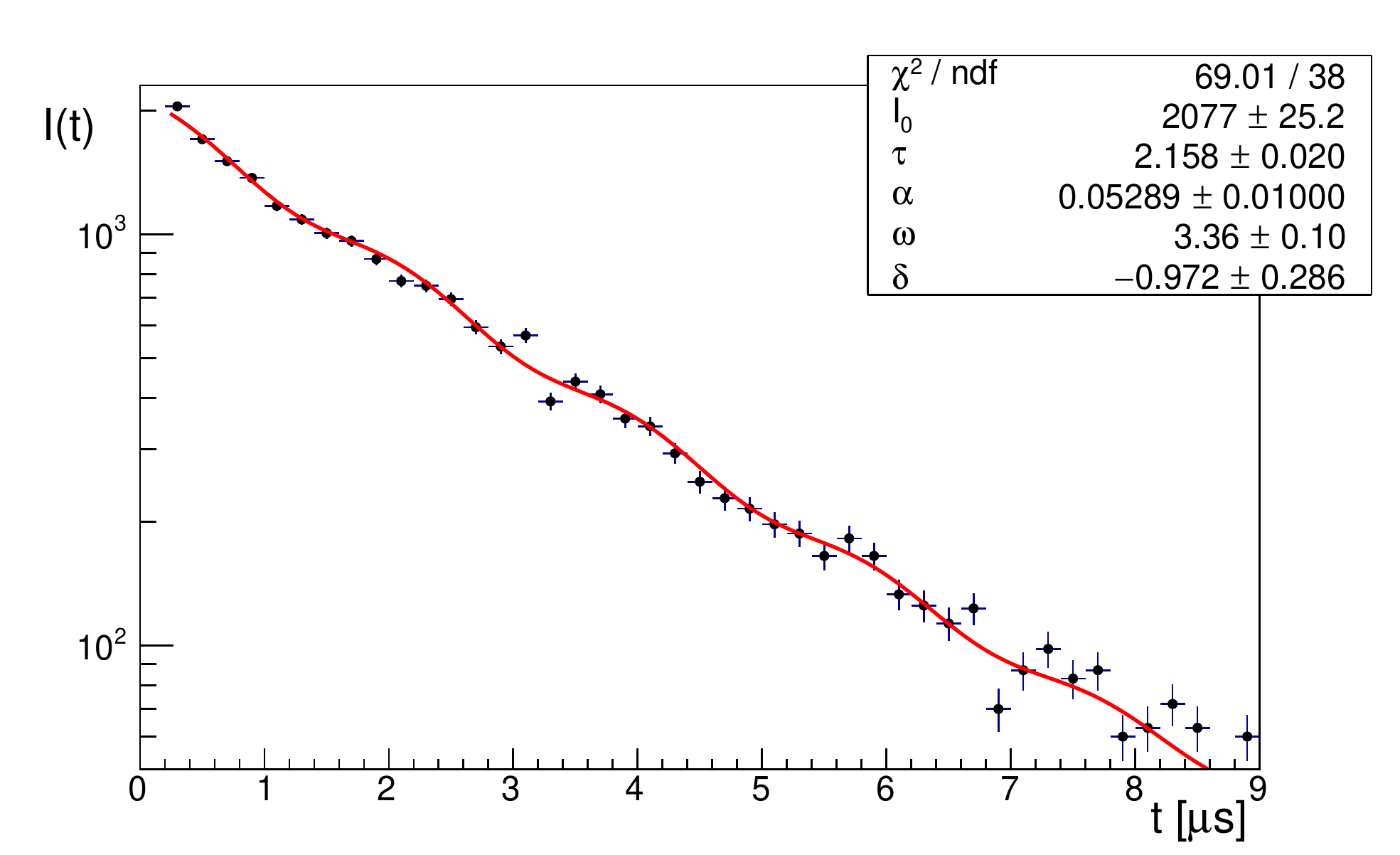}}
{\includegraphics[width=80mm]{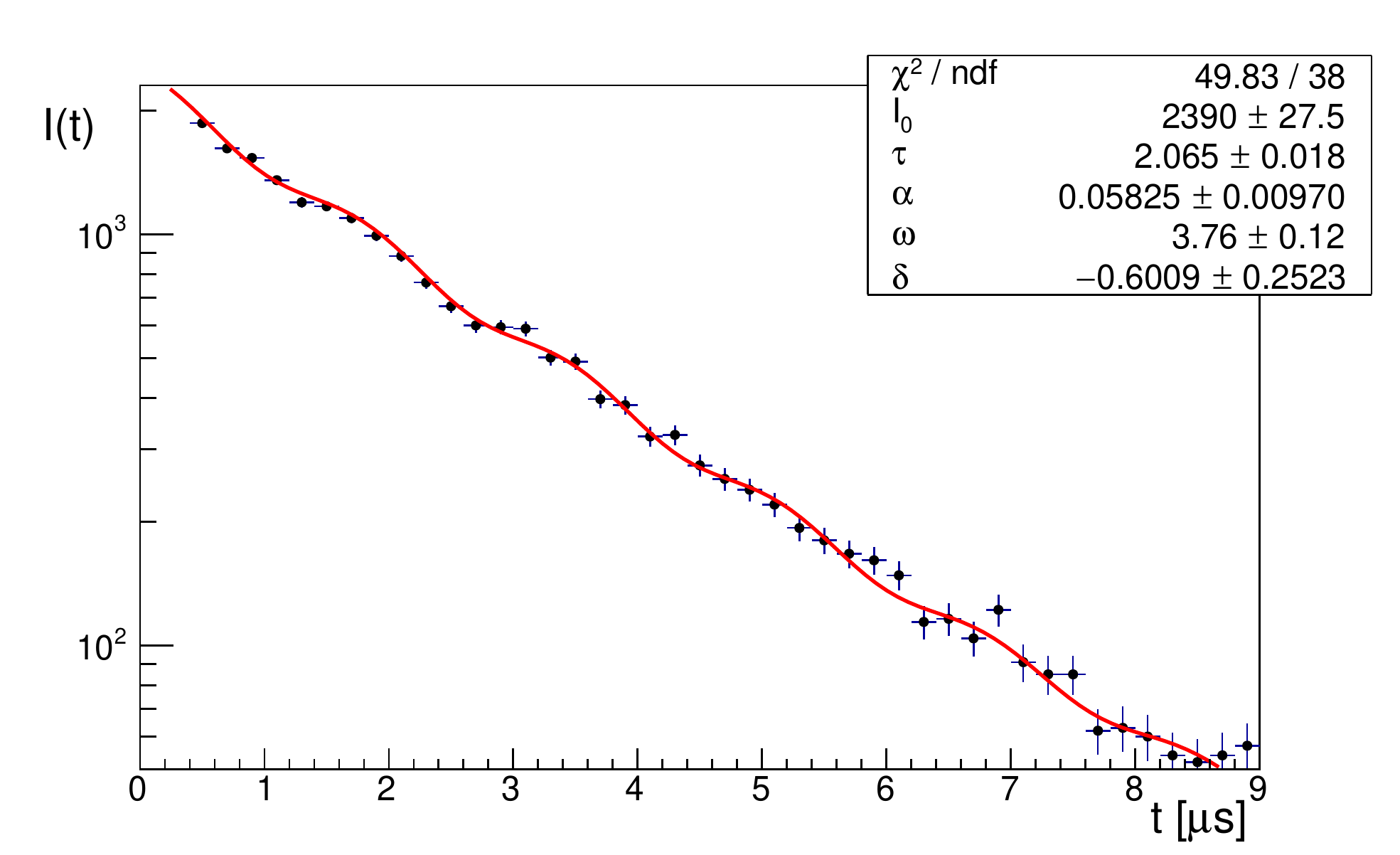}}
\caption{Measurements with the magnetic fields and fits to Eq.~(\ref{eq:I2}), left current of 5 A and right current of 6 A.}
\label{field}
\end{figure}

For the measurements with a magnetic field, within the time window of 10 $\mu$s we obtained 20,936 events (total  10,246,368 registered events), and 22,882 events (total  11,077,886 registered events) with a current of 5 A and 6 A, respectively. The time distributions and fits with the Eq.~(\ref{eq:I2}) for these measurements are shown in Fig.~\ref{field}.

All the fits were performed by using Minuit in the CERN Root software package v6.20.  The fit region was from 0.2 $\mu$s -- 8.8 $\mu$s, which corresponds to approximately four mean muon lifetimes. The constant $C$, which represents the background in the measurement, was not left as a free parameter in the fit but was determined from the data at the large values of $t$. For the measurement of 5 days without the magnetic field, $C$ was approximately 3.5 and for the measurements with the magnetic field, it was approximately 13 for our binning.

The fitting results for the measurements with the magnetic fields and the determination of  $g_{\mu}$ by 
using Eq.~(\ref{eq:omega}) are presented in Table~\ref{fit}. 

\begin{table}[h!]
\centering
\caption{
The results of the fits on the measurements with the magnetic fields
and the determination of $g_{\mu}$.
The uncertainties of the fitted parameters are obtained only by the fitting procedure.
The uncertainty of $g_{\mu}$ is calculated by the error propagation using Eq.
(\ref{eq:omega}) and solved for $g_{\mu}$.}
\begin{ruledtabular}
\begin{tabular}{| c c | c | c | c | c | c | c |}
B [mT] & (I[A])& $\tau$ [$\mu$s] & $\alpha$ & $\omega$ [MHz]& $\delta$ &  $\chi^2$/$ndf$  & $g_{\mu}$ \\ \hline
3.55  $\pm$ 0.15 & (5) & 2.16 $\pm$ 0.02 & 0.05 $\pm$  0.01 &  3.36 $\pm$ 0.10 & -1.0 $\pm$ 0.3 & 69.01/38 &  2.22 $\pm$ 0.11\\ \hline

4.1 $\pm$ 0.2 & (6) & 2.07 $\pm$ 0.02 & 0.06 $\pm$ 0.01 & 3.76 $\pm$ 0.12 & -0.60 $\pm$ 0.25 & 49.83/38 & 2.12 $\pm$ 0.12\\
\end{tabular}
\end{ruledtabular}
\label{fit}
\end{table}

\section{Summary and Discussion}\label{section6}

 We have presented a simple setup for the measurement of the muon magnetic moment suitable for a student laboratory experiment. We exploited the idea of the very effective experiment presented by C. Amsler, \cite{amsler} but we made two significant modifications and improvements.
 
 First, we reduced the detection system to only one scintillation detector compared to three scintillation
  detectors  used in a coincidence/anticoincidence mode in the Amsler experiment. To the best of our knowledge, there are always several scintillation detectors used in various realizations of this experiment as a student laboratory exercise. However, this reduction does not significantly affect the functionality of our system. As also argued in the discussion of the muon lifetime system based on only one scintillation detector, \cite{bosnar} the main source of the background in our approach (if there are no other radioactive sources present) is a second cosmic muon that comes in the selected time window (in our case 10 $\mu$s) after the first registered muon and which is then wrongly interpreted as a positron from the muon decay. The rate of these accidentals, which can be removed by veto in the multiple-detector case, is a decisive factor for the proper functioning of a measuring system comprised of only one scintillation detector. 
  
  Using the statistics of random events for the case of cosmic muons (see, for example,  A.C. Melissinos \emph{et~al.}\cite{melissinos}) and the rate of all incoming cosmic muons that pass through our size of scintillator, 
  it can be shown that the rate of these accidentals in the time window of 10 $\mu$s is indeed not insignificant. However, the overall efficiency of our scintillation detector for the settings used in the measurement was about 12 \%  for the detection of cosmic muons (assuming that there are in total about 200 muons/m$^2\cdot$s). This detection efficiency level also reduces the rate of accidentals, resulting in an expected rate of slightly less than 1 per hour. This estimate is confirmed by the measured total background rate of around 1.5 per hour, which compares favorably to the average of approximately 45 registered events in the time window of 10 $\mu$s per hour. 
  
  We conclude that the rate of muon accidentals is acceptable for the execution of the measurements with our setup. Their contribution
  is taken into account in the off-line analysis  by supposing a flat background in the decay spectra.
  The apparatus is not sensitive enough to quantify the influence of the decays of negative muons, which are caught by nuclei, and this limits the accuracy of the measurements. 
  One of the main sources of the systematic errors also comes from the inhomogeneity in the magnetic field of approx. 5\%.
  
  Second, the maximal simplification in the detector setup enables our other important improvement, that is, the usage of basic and 
  low-cost electronics for the signal processing and data acquisition. Compared to the classical nuclear physics equipment used for such measurements, typically including discriminators, coincidence units, time-to-amplitude converters, and appropriate power crates, our electronics system is more than an order of magnitude lower in cost, and also simpler and more compact, making this experiment readily accessible as a student laboratory experiment. 
  
\begin{acknowledgments}

We would like to thank Ines Mance and  Dra\v{z}en Paunovi\'{c} for their help in the preparation of the cooper coil used in this measurement. This work was supported in part by the Project for the Popularization of Science of the Ministry of Science and Education of Republic of Croatia.
\end{acknowledgments}

\end{document}